\documentclass[12pt]{article}
\usepackage{authblk}
\usepackage{setspace} 
\usepackage{fancyhdr} 
\usepackage[font=small,labelfont=bf]{caption}
\usepackage{booktabs}
\usepackage{enumitem}
\usepackage{tabularx} 
\usepackage{amsmath}  
\usepackage{graphicx, subcaption} 
\usepackage{algorithm}
\usepackage{algorithmicx}
\usepackage{algpseudocode}
\algnewcommand{\LeftComment}[1]{\Statex \(\triangleright\) #1}

\usepackage[margin=1in,letterpaper]{geometry} 
\usepackage{cite} 
\usepackage[final]{hyperref} 
\usepackage [english]{babel}
\usepackage [autostyle, english = american]{csquotes}
\usepackage{amsfonts}
\usepackage{amsmath, amssymb, amsthm}
\usepackage[skip=12pt plus1pt, indent=20pt]{parskip}
\MakeOuterQuote{"}

\usepackage{float}
\hypersetup{
	colorlinks=true,       
	linkcolor=blue,        
	citecolor=green,        
	filecolor=magenta,     
	urlcolor=magenta         
}
\usepackage{physics}
\usepackage{todonotes}
\usepackage{empheq}
 
\newlength\dlf  

\usepackage{gensymb}
\usepackage{mathtools}

\begin{document}\thispagestyle{empty}

\title{The Impact of Approximation on Algorithmic Progress}
\author[1]{Jeffery Li}
\author[1]{Jayson Lynch}
\author[1]{Liva Olina}
\author[1]{Cecilia Chen}
\author[1]{Andrew Lucas}
\author[1]{Neil Thompson}
\affil[1]{MIT CSAIL, Cambridge, United States}

\maketitle

\begin{abstract}
     In nearly every discipline, scientific computations are limited by the cost and speed of computation. For example, the best-known exact algorithms for the canonical Traveling Salesman Problem would take centuries to run on an instance of size 1 million. A natural response to such limits is to try to find new algorithms or to parallelize existing ones, but many algorithms are already at their theoretically-optimal level\cite{liu2021lowerbounds,liu2023optimal} and parallelization is often impossible or prohibitively expensive\cite{tontici2023parallel}. Starting in the 1960's, computer scientists pursued another solution: allowing solutions to have a small amount of error (i.e. approximating them). In this paper, we survey 118 of the most important algorithm problems in computer science, quantifying the gains and tradeoffs from approximation that have been discovered over the history of the field. Overall, only $\approx$20\% of problems have benefited from approximation. However, those with good approximate algorithms can be dramatically faster to compute with little cost to accuracy. For example, a quarter of computationally intractable problems (e.g. those that take exponential time to compute) have polynomial time approximate algorithms. Approximation also increases the number of algorithms that can run in linear time by 23\%, opening up new computational opportunities for those working in the big data regime.  This work also sheds light on what should be expected from progress in AI, where approximation is at the heart of how deep learning works.

 \end{abstract}

One of the main goals of theoretical computer science is to produce fast algorithms. Typically, problems are of the following form: Design an algorithm that, given an input, outputs an answer that \emph{exactly} satisfies some given conditions or optimizes for specific metrics. In practice, given the runtimes of known algorithms, particular problems are only feasibly solvable up to certain input sizes. Take, for example, the canonical Traveling Salesman Problem. While exact algorithms can easily solve small instances, all known exact algorithms for this problem require exponential time or greater to run, making them completely infeasible to run on larger instances. To put this in context, exactly solving just one \emph{instance} of the Traveling Salesman Problem with 85,900 stops took 136 CPU-years \cite{cook2015pursuit}, and even that required using specialized analysis for that specific instance.

In many real-world cases, getting an exact solution is unnecessary. For example, small improvements might have limited economic value, a calculation might have more observational uncertainty than any error introduced algorithmically, or more accurate answers might be infeasible to use (e.g. manufacturing specifications might already be smaller than the precisions of the tools being used to build them). In such cases, an approximation algorithm could achieve a solution much faster with no downside. In other cases, there might be a tradeoff between accuracy and runtime. For the Traveling Salesman Problem, a much more efficient approximate algorithm exists that runs in cubic time ($n^3$) and gets within a factor of $1.5$ of the optimal solution. This can easily approximate instances with up to 100,000 stops in minutes, but introduces some error. 

This approximability question has important implications for AI. A neural network of a given size can only do a specific number of calculations (or a multiple of this if thinking / chain of thought is used). But results from complexity theory, such as the Time Hierarchy Theorem\cite{hartmanis1965computational}, show that problems cannot be solved \textit{exactly} with such fixed amounts of compute. Thus, necessarily, neural networks are approximating any such functions. Further, inapproximability results reveal that some problems must have large errors when approximated. It is thus of central interest which functions can be approximated, because it provides a limit to the functions that neural networks should reliably be able to solve.

Approximation algorithms can not only open up new computational opportunities by allowing larger problem sizes to be tackled, but also dramatically decrease the computation needed for an existing problem. Often such speedups result in proportional reductions in financial costs and the energy and carbon footprint of the calculations. But just because such dramatic improvements are possible, doesn't mean that they are common enough to matter for the improvement of computing usage more generally. To understand that question, we perform the first large-scale study of approximation algorithms.

\subsection*{Prior Work}

Leiserson et al. \cite{leiserson2020moore} analyzed the state of improvements in computing after the end of Moore's law and argued that progress will come in three main areas: hardware architecture, software, and algorithms. In particular, solving or approximating an algorithm problem with faster algorithms will improve our effective computing power. As such, it is important to review the state of algorithmic improvements and see where additional efficiency gains are possible.

The first paper to address the rate of algorithmic progress on a large scale is due to Sherry and Thompson in 2021 \cite{sherry2021fast}. They analyzed improvements in runtimes for 113 algorithm problems, finding that 14\% of problems have seen dramatic improvements, 30-43\% of problems have seen improvements at least as significant as Moore's Law, but around half of the problems see minimal improvement. Rome et al. \cite{rome2023space,rome2023paper} performed similar analysis for auxiliary space complexity on this set of problems, finding that an even greater percentage (80\%) of problems see minimal improvement for memory usage. Together, these suggest that finding asymptotic improvements to exact algorithms is challenging.

To study why this might be the case, Liu et al. \cite{liu2021lowerbounds,liu2023optimal} analyzed algorithmic lower bounds. They found that around half of the problems in their dataset already have asymptotically optimal algorithms and thus no further asymptotic improvements are possible.

Recently, Tontici \cite{tontici2023parallel} analyzed parallel algorithms, which utilize multiple resources such as processors to run steps in parallel. They found that over half of the problems in their dataset had at least one parallel algorithm, showing that a wide range of problems benefit from parallelism. Relative speedups typically range from around a factor of 10 or 100 for personal computers to around $10^6$ for supercomputers, depending on the number of processors available. This gain may not be enough to tackle computationally challenging problems feasibly. Indeed, by contrast, approximation algorithms can have gains many orders of magnitude larger. In addition, it is important to note that running steps in parallel does not reduce the overall amount of computation required to solve a problem; the computation only gets divided among the processors.

These papers analyze \emph{exact} algorithms; they exclude approximation algorithms from their analyses. To our knowledge, no preceding or following paper seems to have analyzed approximation algorithms on a large scale; previous sources like \cite{srinivasan1999approximation,books/cu/MotwaniR95,vazirani2001approximation,williamson2011design} focus only on specific problems or techniques.

\section*{Results}\label{ch4:approx}

\subsection*{Overview}

Approximation algorithms originated in the 1600s with many root-finding methods, whose goal is to determine a \emph{root} of a real-valued continuous function $f(x)$, or a real number $r$ such that $f(r)=0$. One approximation algorithm for root finding is the bisection method. Given two initial values $a<b$ with $f(a)$ and $f(b)$ of opposite signs, the bisection method repeatedly halves the interval $[a,b]$ and keeps the subinterval where the sign change remains. Each halving shrinks the interval exponentially until its length is at most $\varepsilon$. The midpoint of that interval is then within $\frac{\varepsilon}{2}$ of the true root. This shows that the bisection method is an approximation algorithm with \emph{additive error}.

Additive error is not the only type of error approximation algorithms consider; many approximation algorithms, especially those introduced after the 1960s, have \emph{multiplicative error}, where the error term describes the worst possible ratio between the values of the algorithm's answer and the correct/optimal answer. In other words, an approximation algorithm for a minimization problem has a multiplicative error of $\alpha$ if the output $\tilde{k}$ it returns satisfies $k\leq \tilde{k}\leq \alpha k$, where $k$ is the true answer. Approximation algorithms can also have \emph{additive-multiplicative error}, meaning that they produce an answer that has both additive error and multiplicative error components.

\begin{figure}[!ht]
    \centering
    \includegraphics[width=\textwidth]{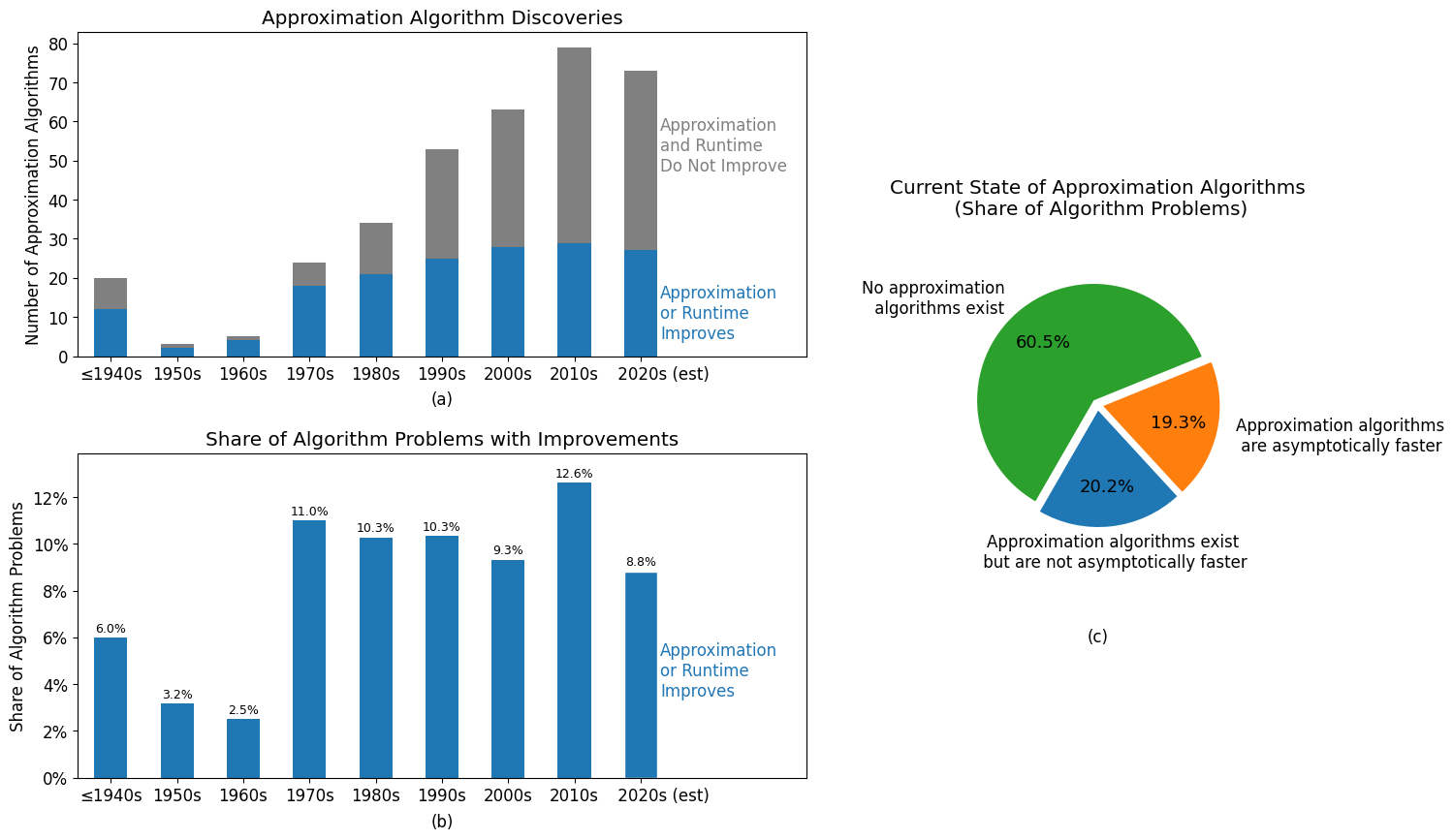}
    \caption{Historical progress and current state of approximation algorithms. 2020s estimates assume that future years values will (proportionally) follow a trajectory similar to the 2010s.}
    \label{fig:fig1}
\end{figure}

Since the introduction of combinatorial approximation algorithms in the 1960s, the field of approximation algorithms has grown, as shown in Figure \ref{fig:fig1}. Improvements from approximation algorithms are not localized to one or two algorithm problems; for each decade since the 1970s, around 8-12\% of the algorithm problems in our dataset see these improvements. In total, approximation algorithms have provided significant asymptotic runtime improvements for around one-fifth of the algorithm problems in our dataset. These are large proportions, considering that only 40\% of the algorithm problems in our dataset have approximation algorithms, and highlight how approximation can provide substantial efficiency gains for a sizable group of problems. 

\subsection*{Algorithmic Improvements in Theory}\label{sec:approx_improve_theory}

In this section, we analyze the theoretical speedup of approximation algorithms at different levels of error.

\begin{figure}[!ht]
    \centering
    \includegraphics[width=0.9\textwidth]{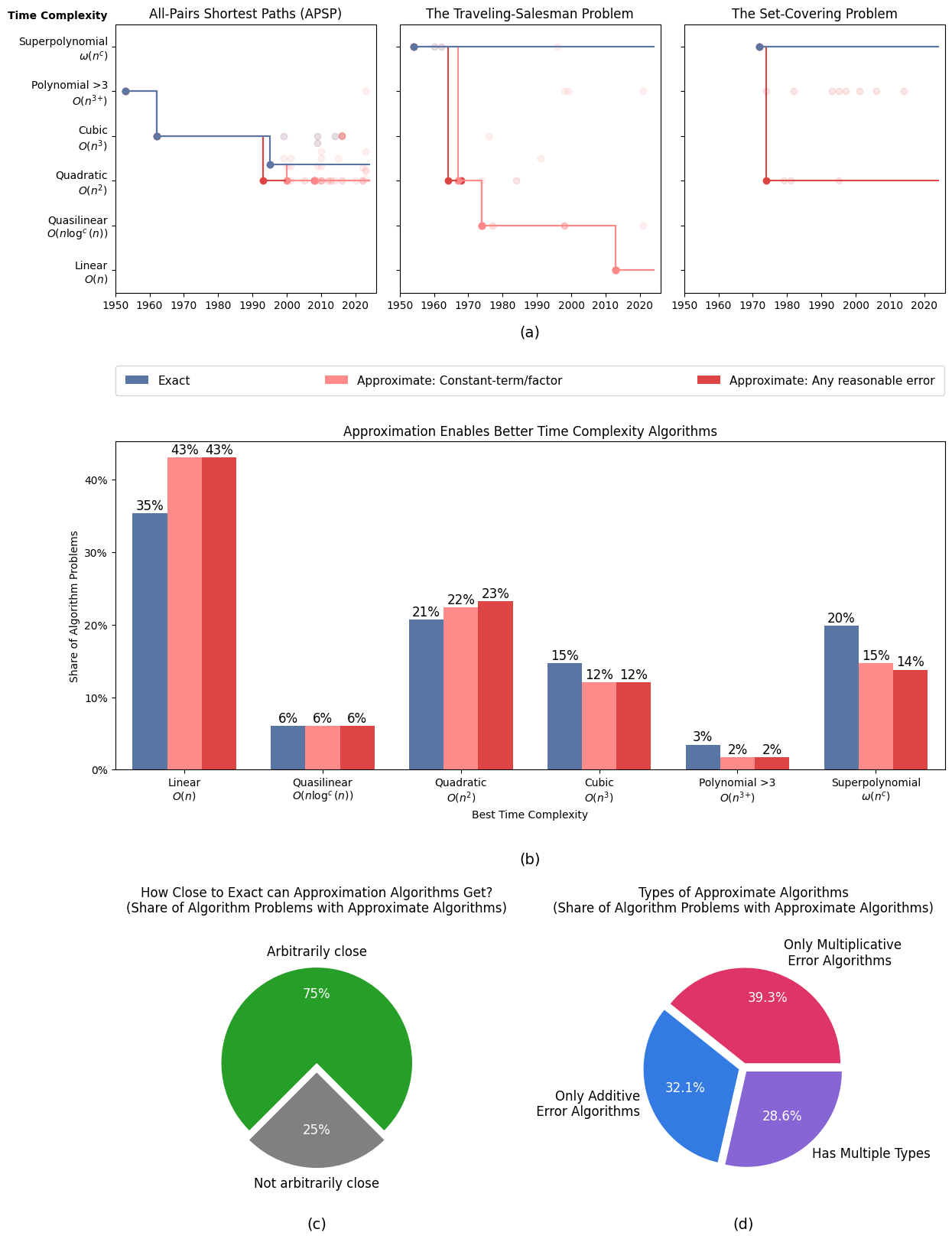}
    \caption{Theoretical improvements and characteristics (arbitrarily closeness, types of error) of approximation algorithms. For (a), faded dots indicate algorithms that do not improve the best runtime. Dots and lines between tick marks represent intermediate values. These plots are more specifically for Undirected, Unweighted All-Pairs Shortest Paths, and for the Geometric Traveling Salesman Problem.}
    \label{fig:fig2}
\end{figure}

We first consider specific problems. Figure \ref{fig:fig2}(a) shows how the best theoretical runtime has evolved over the years for three problems: \emph{All-Pairs Shortest Paths}, the \emph{Traveling Salesman Problem}, and the \emph{Set Cover Problem}. We consider the following three levels of error: no error, constant-term or better error\footnote{We treat approximation error parameters as constants, as they are typically set to small values and are thus bounded.}, and any reasonable error\footnote{We avoid trivial algorithms, like those that ignore the input and simply output an arbitrary answer.}.

In \emph{All-Pairs Shortest Paths}, the goal is to determine, for all pairs of vertices in a given graph with $V$ vertices, the shortest path between the two vertices. Here, we specifically study the variant where all edges are unweighted and undirected. All-Pairs Shortest Paths is a problem with many cubic-time exact algorithms\footnote{The only subcubic-time exact algorithm for All-Pairs Shortest Paths in our figure uses fast matrix multiplication methods, many of which are impractical (see page 2 of \cite{legall2012matmul}).}. In addition, it has faster approximation algorithms, many of which are subcubic-time, including one introduced in 1999 by Aingworth et al. \cite{aingworth1999fast} that has an additive error of $2$ and runs in time $O(V^{2.5}\sqrt{\log V})$. This shows how approximation algorithms are important even in the polynomial-time regime, as they can help shave off polynomial factors in runtime with small costs in accuracy. Nearly quadratic-time algorithms with worse approximation factors also exist \cite{baswana2009all}. However, the runtime gap between exact and approximation algorithms is small (i.e., linear or smaller), which is the case for several algorithm problems.

In the \emph{Traveling Salesman Problem}, the goal is to determine the minimum edge weight cycle which visits all $V$ vertices of a given weighted graph exactly once. Here, we specifically study the geometric variant\footnote{In the geometric variant, the graph is embedded in Euclidean space, and edge weights are the Euclidean distance between the two vertices.}. This problem is $NP$-hard, meaning no exact polynomial-time algorithms exist unless $P=NP$, and the best known exact algorithms have exponential runtime. However, researchers have done significant work to reduce the runtime down to linear in $V$, albeit with large constant factors in the runtime.

In the \emph{Set Cover Problem}, the goal is, given a collection $S$ of sets, to determine the smallest sub-collection $C$ of sets such that the union of the sets in $C$ equals the union of the sets in $S$. Like the Traveling Salesman Problem, the Set Cover Problem is an $NP$-hard problem, but it also has a conditional $(1-o(1))\ln n$-approximation lower bound \cite{dinur2013setcover}, so even constant-factor approximation algorithms will likely have superpolynomial runtime. However, we see quadratic-time approximation algorithms if we allow for $\log n$ error.

Figure \ref{fig:fig2}(b) shows the distribution of all algorithm problems based on the best theoretical runtime of any algorithm at the three levels of error. There is a noticeable shift in the pink/red bars towards lower runtimes, reflecting the power of approximate algorithms to produce asymptotic improvements. These improvements take several forms. Some problems in our dataset are $NP$-hard, so they fall in the superpolynomial class when only considering exact algorithms. Between one-fourth and one-third of these can be improved to polynomial time through approximation. Even more striking is the significant increase in the "linear" class, which jumps from 35\% to 43\% with approximation. Linear algorithms are particularly important for big data analyses, making approximation disproportionately important in those areas. Interestingly, there is little change if we allow for more error (pink vs red bars); only one problem, the \emph{Set Cover Problem}, changes runtime classes.

One common way approximation algorithms trade off between speed and accuracy is through an \emph{approximation error parameter} $\varepsilon$. Such a parameter allows the algorithm user to choose how close they want to be to the correct/optimal answer based on how long they want to run the algorithm. For many (but not all) such algorithms, decreasing $\varepsilon$ yields arbitrarily high accuracy at the cost of added runtime. For example, in the bisection method discussed earlier, $\varepsilon$ directly controls the guaranteed accuracy of the output: smaller $\varepsilon$ improves accuracy guarantees but increases runtime as more iterations are needed before the interval length drops below $\varepsilon$.

Figure \ref{fig:fig2}(c) shows, among algorithm problems that have approximation algorithms, the proportion that have an approximation algorithm that can achieve arbitrarily small error through the approximation parameter $\varepsilon$. Interestingly, a majority (75\%) of these algorithm problems fall into this category. Many of these problems are either numerical analysis-based problems with iterative algorithms with additive $\varepsilon$ error, or NP-hard problems with \textbf{polynomial-time approximation schemes}. Polynomial-time approximation schemes have $1+\varepsilon$ multiplicative error and polynomial runtimes in the input size for every fixed value of $\varepsilon$, like $O(n^{1+1/\varepsilon})$.

We also analyzed the types of error introduced by approximation algorithms. Different problem types are conducive to different error types; numerical analysis-based problems will usually have additive-error approximation algorithms, while combinatorial problems will usually have multiplicative-error approximation algorithms. Figure \ref{fig:fig2}(d) splits the algorithm problems with approximation algorithms based on what types of error their algorithms have. Multiplicative error only is the most common at roughly 40\%, while roughly 32\% admit only additive error. For both types of error, the error terms are usually either small constants or parameterized, suggesting that error guarantees usually do not hinder practical use of these approximation algorithms. However, around 15\% of the algorithms in our dataset have large errors, exceeding an additive term or multiplicative factor of $10$ for large problem sizes.
Such error sizes can be significant; for example, when solving equations involving low-energy systems or slow reactions, where solutions are small in magnitude, having an additive error of $10$ provides large uncertainty, and in a real-world instance of the Traveling Salesman Problem, being a multiplicative factor of $2$ off translates to a traveled route that is twice as long, leading to a significant amount of wasted resources.
This suggests that some approximation algorithms may be impractical due to weak approximation guarantees.

\subsection*{Quantitative Impact of Algorithmic Improvements}\label{sec:approx_improve_practice}

How do theoretical runtime improvements translate to speedups for specific problem sizes? Here, we analyze the speedup factors for particular problem sizes. Since we ignore constant factors in runtimes, these figures are rough estimates, but Sherry and Thompson \cite{sherry2021fast} showed that constant factors typically have small effects on these types of calculations. 

\begin{figure}[!ht]
    \centering
    \includegraphics[width=\textwidth]{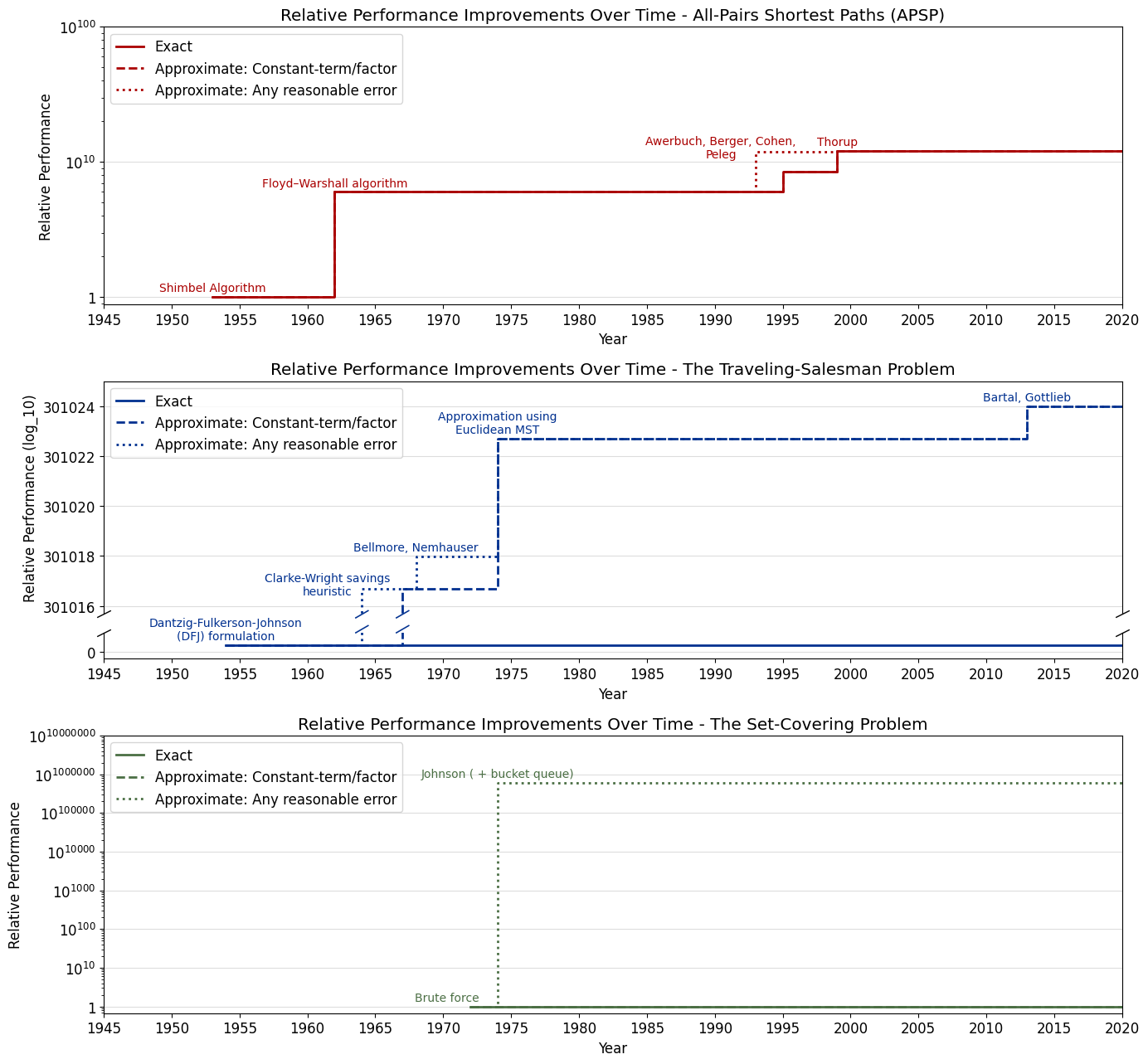}
    \caption{Relative improvements over time for three different problems (All-Pairs Shortest Paths, Traveling Salesman Problem, and Set Cover Problem) at three different levels of error, using problem size $n=10^6$.}
    \label{fig:fig3}
\end{figure}

First, we revisit the three problems from the previous subsection: \emph{All-Pairs Shortest Paths}, \emph{Traveling Salesman Problem}, and \emph{Set Cover Problem}. Figure \ref{fig:fig3} shows, for each problem, the relative performance improvements over time at the three levels of error discussed in the previous subsection, which are "no error," "constant-term or better error," and "any reasonable error," using a problem size of $n = 10^6$. 

We see the two general types of improvements that algorithm problems get from approximation algorithms. On one hand, we have problems like All-Pairs Shortest Paths. These problems see small improvements due to slight improvements to the polynomial exponent in the runtime, that occasionally get matched by later improvements from exact algorithms.
On the other hand, we have problems like the Traveling Salesman Problem and the Set Cover Problem, which see enormous improvements corresponding to jumps from exponential-time exact algorithms to the first polynomial-time approximation algorithm. The Traveling Salesman Problem later sees several smaller jumps due to improved polynomial exponents.

To understand the importance of approximation across all problems, we convert the overall performance improvement into an amortized annual improvement rate. We use a modified version of the \textbf{compound growth rate} \cite{sherry2021fast}: \[ \text{CAGR} = (r_e/r_{a, \varepsilon})^{1/t}-1,\] where $t$ is the number of years since the first algorithm for the problem and $r_e$ and $r_{a, \varepsilon}$ are the number of steps associated with the asymptotic runtimes of the best exact algorithm and the best approximation algorithm with error term at most $\varepsilon$, respectively.

\begin{figure}[!ht]
    \centering
    \includegraphics[width=\textwidth]{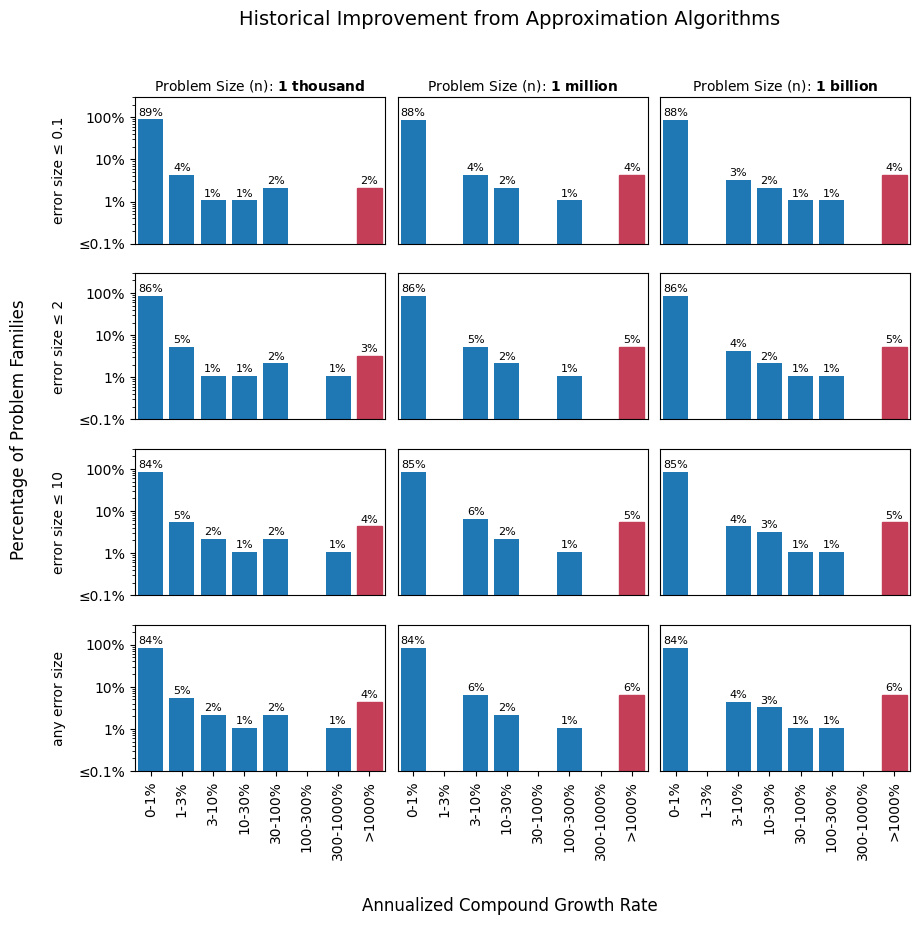}
    \caption{Distributions of the annualized improvements the best approximation algorithms achieve over the best exact algorithms, measured from the year of the first algorithm for each problem, across different problem sizes ($n$) and error tolerances (the answer returned by the algorithm is within this additive term or multiplicative factor of the optimal).}
    \label{fig:fig4}
\end{figure}

Figure \ref{fig:fig4} shows the distribution of compound growth rates across all 118 algorithm problems for 12 different combinations of problem size $n$ ($10^3$, $10^6$, and $10^9$) and error tolerance $\varepsilon$ ($0.1$, $2$, $10$, and any).

We see that improvements are heterogeneous across algorithm problems. Around 85-90\% of problems see no noticeable speedups regardless of problem size and error tolerance, showing that approximation is not compatible with many problems. On the other hand, several problems observe enormous speedups, with growth rates over 1000\% per year. Many of these problems are $NP$-hard problems with polynomial-time approximation algorithms, where the drop from exponential to polynomial translates to an enormous improvement.

Problem size and error tolerance have small effects on the overall distribution of compound growth rates. Around 10\% of problems change categories as problem size or error tolerance increase, with each individual category changing by at most 5\%. The main changes occur when the problem size increases from $10^3$ to $10^6$ and when the error tolerance increases from $2$ to $10$. These changes are caused by algorithms introducing asymptotic runtime improvements, which are scale-dependent, at the cost of larger approximation errors. Besides these small changes, the distributions are similarly shaped across all combinations of problem size and error tolerance considered. Based on how we categorized the growth rates, this suggests that the scale-dependent improvements typically fall under one of two extremes (either negligible or enormous), with very few being moderately-sized.

\section*{Discussion}\label{ch6:conclusion}

Algorithmic progress has come in waves: exact serial algorithms peaked in the 1970s, and parallel algorithms peaked in the 1980s \cite{tontici2023parallel}, while approximation algorithms continue to grow. Parallelization is becoming increasingly important, but with the current chips available, parallelization rarely gives more than a millionfold factor in speedup over serial algorithms, even on supercomputers \cite{tontici2023parallel}, which is significantly smaller than the speedups produced by reducing the running time of an algorithm from exponential to a small polynomial. For several problems, approximation algorithms achieve such speedups with only slight costs to accuracy.

Approximation is important in two key ways: (i) it makes problems whose best known exact algorithms run in exponential time accessible through polynomial-time approximation algorithms, and (ii) it enables more linear-time algorithms, which is key for very large-scale big data analyses. Both expand the range of feasible computational analyses, but only a small set of problems benefit substantially, showing that approximation is not broadly compatible across algorithm problems.

This work also provides more insight into the current explosion of work in AI and machine learning. Inherently, deep learning produces approximate answers (indeed, it is a universal function approximator \cite{hornik1989multilayer}), so one might hope that it could learn good approximations across a wide range of applications. If we believe that traditional algorithm problems are representative of a large portion of problems we may want to solve, or are critical sub-steps to them, then our analysis suggests skepticism of this hope for deep learning. If few traditional algorithm problems have approximate algorithms with performance guarantees, many AI-learned functions will also likely lack these guarantees. As such, these AI-based approximations could be intrinsically less robust than the algorithms they are replacing.

Finally, there is substantial work on when algorithmic problems can be approximated. Future work could involve a systemic study of inapproximability results and understanding how close current approximation algorithms are to being optimal, akin to Liu's work on exact algorithms\cite{liu2021lowerbounds, liu2023optimal}. This could further support the view that we should not expect machine learning algorithms to provide significantly more efficient and applicable approximations across all problems of interest.

\section*{Methods}\label{ch3:methods}

In this section, we discuss our approach to collecting and analyzing the data used to obtain the results in previous sections. We will cover the scope and process of our data collection and how our data is processed and used to produce our analysis and results.

\subsection*{Scope}\label{sec:scope}

To allow for comparisons with algorithmic gains from exact serial algorithms and parallel algorithms, we consider the same set of algorithm problems used in previous analyses by Sherry and Thompson \cite{sherry2021fast}, Rome \cite{rome2023space,rome2023paper}, and Tontici \cite{tontici2023parallel}. We define an \textbf{algorithm problem} as a collection of various formulations of the same general task. These formulations have the same structure, but they vary in restrictions on input values, format of the output, or the metric used to judge the output. We call these \textbf{problem variations}. We made sure to account for the fact that some algorithms may only solve or be effective for specific variations.

We use the same set of \textbf{118 algorithm problems} from Rome \cite{rome2023space,rome2023paper} for our analysis, and many of our figures consider all 118 algorithm problems. However, many algorithm problems currently see no benefits from approximation algorithms. Out of the 118 algorithm problems, 60 do not have any approximation algorithms to collect, either because it is difficult to define a "metric" on the outputs (particularly for decision, or yes/no, problems like the \emph{Graph Isomorphism Problem}) or because there hasn't been sufficient interest for designing approximation algorithms for these problems. Once we filter out irrelevant approximation algorithms, particularly those whose asymptotic runtime or approximation factor haven't been analyzed or those that don't solve a relevant variant, 30 more algorithm problems do not have any relevant approximation algorithms. As such, we end up with \textbf{28} algorithm problems that have relevant approximation algorithms.

\subsection*{Data Collection}\label{sec:datacollection}

We extract about 80 approximation algorithms from the database of algorithms collected for previous works \cite{rome2023space,rome2023paper}. For the rest of the approximation algorithms dataset, we collect approximation algorithms by looking through papers, mostly through Google Scholar and Wikipedia, and using references in these papers. We have 332 approximation algorithms making up our analyzable dataset.

The most relevant information we extract for each approximation algorithm is:
\begin{itemize}
    \item The algorithm problem and variation
    \item The authors and year of publication of the paper
    \item The worst-case time complexity
    \item Any parameter definitions (like number of vertices or edges in a graph or number of elements in the input)
    \item The model of computation (typically the word RAM or real RAM models of computation)
    \item Approximation error term, error type (additive, multiplicative, additive-multiplicative), whether or not the error is parameterized,
    and a brief description of the error.
\end{itemize}

As is standard in analysis of algorithms, we use \emph{asymptotic notation} for time complexities, ignoring constant factors and lower-order terms. For example, an algorithm that solves a problem in $2n^3+3n$ steps has a time complexity of $O(n^3)$.

For many problems, the time complexity is commonly expressed in terms of the input size $n$; however, this is not always the case. For example, many algorithms for graph problems have time complexities written in terms of $|V|$ when the input size is $|V|+|E|$, where $|V|$ and $|E|$ are the numbers of vertices and edges, respectively. We follow previous papers \cite{rome2023space,rome2023paper,tontici2023parallel} in standardizing time complexities in terms of the problem size, so in the case of algorithms for graph problems, we use $n = |V|+|E|$.

\subsection*{Data Processing}\label{sec:dataprocessing}

We create a classification scheme for time 
complexities and approximation errors to have a standardized way of analyzing improvements in time complexity or approximation. We also consider intermediate values between these classifications for algorithms whose asymptotic runtimes fall in between two classes.

We group algorithmic running times into eight time complexity classes; these classes can be found in Table~\ref{tab:Time Complexity}. For approximation factors, as there are both additive and multiplicative errors, we use two different classifications. These classes are given in Tables~\ref{tab:Additive Error Classes} and \ref{tab:Multiplicative Error Classes}.

\begin{table}[!ht]
\centering
\begin{tabular}{|l|l|}
\hline
\multicolumn{2}{|c|}{\textbf{Time Complexity Classes}} \\
\hline
Constant & $O(1)$ \\
Polylogarithmic & $O(\log^c n)$ for some constant $c>0$ \\
Linear & $O(n)$ \\
Quasilinear & $O(n \log^c n)$ for some constant $c>0$ \\
Quadratic & $O(n^2)$ \\
Cubic & $O(n^3)$ \\
Supercubic but still polynomial & $O(n^c)$ for some constant $c>3$ \\
Superpolynomial & $\omega(n^c)$ for any constant $c>0$ \\
\hline
\end{tabular}
\caption{Grouped asymptotic worst case time complexities.}
\label{tab:Time Complexity}
\end{table}

\begin{table}[!ht]
\centering
\begin{tabular}{|l|l|}
\hline
\multicolumn{2}{|c|}{\textbf{Additive Error Classes}} \\
\hline
Sub-constant & $o(1)$ (for example, $1/n^c$ for any $c>0$) \\
Constant & $O(1)$ \\
Logarithmic & $O(\log n)$ \\
Polylogarithmic (exponent $>$ 1) & $O(\log^c n)$ for some constant $c>1$ \\
Square root & $O(\sqrt{n})$ \\
Linear & $O(n)$ \\
Quadratic or greater & $\Omega(n^2)$ \\
\hline
\end{tabular}
\caption{Classes of additive approximation errors.}
\label{tab:Additive Error Classes}
\end{table}

\begin{table}[!ht]
\centering
\begin{tabular}{|l|l|}
\hline
\multicolumn{2}{|c|}{\textbf{Multiplicative Error Classes}} \\
\hline
Approaching 1 & $1+o(1)$ (for example, $1+(1/n^c)$ for any $c>0$) \\
Arbitrarily small & $1+\varepsilon$ \\
Factor of 2 & $2$ \\
Constant $>$ 2 & $O(1)$ greater than $2$ \\
Double logarithmic & $O(\log \log n)$ \\
Logarithmic & $O(\log n)$ \\
Polylogarithmic (exponent $>$ 1) & $O(\log^c n)$ for some constant $c>1$ \\
Square root & $O(\sqrt{n})$ \\
Linear & $O(n)$ \\
Superlinear & $\omega(n)$ \\
\hline
\end{tabular}
\caption{Classes of multiplicative approximation errors.}
\label{tab:Multiplicative Error Classes}
\end{table}

\bibliographystyle{bibstyle}
\bibliography{biblio}

@article{liu2021lowerbounds,
  title={A {M}etastudy of {A}lgorithm {L}ower {B}ounds},
  author={Liu, Emily},
  year={2021},
  journal={Master’s thesis, Massachusetts Institute of Technology}
}

@article{rome2023space,
  title={The {S}pace {R}ace: {P}rogress in {A}lgorithm {S}pace {C}omplexity},
  author={Rome, Hayden},
  year={2023},
  journal={Master’s thesis, Massachusetts Institute of Technology}
}

@article{tontici2023parallel,
  title={Progress in {P}arallel {A}lgorithms},
  author={Tontici, Damian},
  year={2024},
  journal={Master’s thesis, Massachusetts Institute of Technology}
}

@article{sherry2021fast,
  title={How {F}ast do {A}lgorithms {I}mprove? [{P}oint of {V}iew]},
  author={Sherry, Yash and Thompson, Neil C},
  journal={Proceedings of the IEEE},
  volume={109},
  number={11},
  pages={1768--1777},
  year={2021},
  publisher={IEEE}
}

@article{srinivasan1999approximation,
  title={Approximation algorithms via randomized rounding: a survey},
  author={Srinivasan, Aravind},
  journal={Series in Advanced Topics in Mathematics, Polish Scientific Publishers PWN},
  pages={9--71},
  year={1999}
}

@article{dinur2013setcover,
  author       = {Irit Dinur and
                  David Steurer},
  title        = {Analytical {A}pproach to {P}arallel {R}epetition},
  journal      = {CoRR},
  volume       = {abs/1305.1979},
  year         = {2013},
  url          = {http://arxiv.org/abs/1305.1979},
  eprinttype    = {arXiv},
  eprint       = {1305.1979},
  bibsource    = {dblp computer science bibliography, https://dblp.org}
}

@article{liu2023optimal,
  title={How {C}lose are {A}lgorithms to being {O}ptimal?},
  author={Liu, Emily and Sherry, Yash and Kuszmaul, William  and Lynch, Jayson and Thompson, Neil C.},
  year={2023},
  journal = {Working Paper}
}

@article{rome2023paper,
  title={How {F}ast are {A}lgorithms {R}educing the {D}emands on {M}emory?  {A} {S}urvey of {P}rogress in {S}pace {C}omplexity},
  author={Rome, Hayden and Lynch, Jayson and Li, Jeffery and Falor, Chirag and Thompson, Neil C.},
  year={2024},
  journal = {Working Paper}
}

@article{leiserson2020moore,
author = {Charles E. Leiserson  and Neil C. Thompson  and Joel S. Emer  and Bradley C. Kuszmaul  and Butler W. Lampson  and Daniel Sanchez  and Tao B. Schardl },
title = {There’s plenty of room at the {T}op: {W}hat will drive computer performance after {M}oore’s law?},
journal = {Science},
volume = {368},
number = {6495},
pages = {eaam9744},
year = {2020},
doi = {10.1126/science.aam9744},
URL = {https://www.science.org/doi/abs/10.1126/science.aam9744},
eprint = {https://www.science.org/doi/pdf/10.1126/science.aam9744}}

@article{legall2012matmul,
  author       = {Fran{\c{c}}ois Le Gall},
  title        = {Faster {A}lgorithms for {R}ectangular {M}atrix {M}ultiplication},
  journal      = {CoRR},
  volume       = {abs/1204.1111},
  year         = {2012},
  url          = {http://arxiv.org/abs/1204.1111},
  eprinttype    = {arXiv},
  eprint       = {1204.1111},
  bibsource    = {dblp computer science bibliography, https://dblp.org}
}

@book{cook2015pursuit,
  title={In Pursuit of the Traveling Salesman: Mathematics at the Limits of Computation},
  author={Cook, William J},
  year={2015},
  publisher={Princeton University Press}
}

@article{aingworth1999fast,
  title={Fast {E}stimation of {D}iameter and {S}hortest {P}aths ({W}ithout {M}atrix {M}ultiplication)},
  author={Aingworth, Donald and Chekuri, Chandra and Indyk, Piotr and Motwani, Rajeev},
  journal={SIAM Journal on Computing},
  volume={28},
  number={4},
  pages={1167--1181},
  year={1999},
  publisher={SIAM}
}

@book{books/cu/MotwaniR95,
  author = {Motwani, Rajeev and Raghavan, Prabhakar},
  ee = {https://doi.org/10.1017/cbo9780511814075},
  isbn = {9780511814075},
  publisher = {Cambridge University Press},
  title = {Randomized Algorithms.},
  year = 1995
}

@book{vazirani2001approximation,
  title={Approximation Algorithms},
  author={Vazirani, Vijay V},
  volume={1},
  publisher={Springer},
  address={Berlin},
  year={2001}
}

@book{williamson2011design,
  title={The Design of Approximation Algorithms},
  author={Williamson, David P and Shmoys, David B},
  year={2011},
  publisher={Cambridge university press}
}

@article{hornik1989multilayer,
  title={Multilayer feedforward networks are universal approximators},
  author={Hornik, Kurt and Stinchcombe, Maxwell and White, Halbert},
  journal={Neural networks},
  volume={2},
  number={5},
  pages={359--366},
  year={1989},
  publisher={Elsevier}
}

@article{baswana2009all,
  title={All-pairs nearly 2-approximate shortest paths in $O(n^2 \text{ polylog } n)$ time},
  author={Baswana, Surender and Goyal, Vishrut and Sen, Sandeep},
  journal={Theoretical Computer Science},
  volume={410},
  number={1},
  pages={84--93},
  year={2009},
  publisher={Elsevier}
}

@article{hartmanis1965computational,
  title={On the computational complexity of algorithms},
  author={Hartmanis, Juris and Stearns, Richard E},
  journal={Transactions of the American Mathematical Society},
  volume={117},
  pages={285--306},
  year={1965},
  publisher={JSTOR}
}

\end{document}